\documentclass{aa}  
\usepackage{graphicx}
\usepackage{txfonts}
\begin{document} 

\title{The Detection of Earth-mass Planets around Active Stars: }
\subtitle{The Mass of Kepler-78b}

\author{A. P. Hatzes}

  \institute{Th\"uringer Landessternwarte Tautenburg,
                Sternwarte 5, D-07778 Tautenburg, Germany}

\date{Received 17 April 2014 / Accepted 28 June 2014}

 
\abstract{Kepler-78b is a transiting Earth-mass planet in an 8.5 hr orbit discovered by the Kepler Space Mission.
 We performed an
analysis of the published  radial velocity measurements for Kepler-78 in order to derive
a refined measurement for the planet mass. Kepler-78 is an active star and radial velocity
variations due to activity were removed using a Floating Chunk Offset (FCO) method where an 
orbital solution was made
to the data by allowing the velocity offsets  of individual nights to vary. {We show that
if we had no a priori knowledge of the transit period 
the FCO method used as a periodogram  would still have detected 
Kepler-78b in the radial velocity data. It can thus 
be effective
at finding unknown short-period signals in the presence of significant activity noise. }
Using the FCO method {while
keeping the ephemeris and orbital phase fixed to the photometric values} and using only data
from nights where 6-10 measurements were taken results in a $K$-amplitude of 1.34 $\pm$ 0.25 m\,s$^{-1}$,
a planet mass of 1.31 $\pm$ 0.24 $M_\oplus$, and a planet density of 
$\rho$ = $4.5_{-2.0}^{+2.2}$ g\,cm$^{-3}$. {Allowing the orbital phase to be a
free parameter reproduces the transit phase  to within the uncertainty.} 
The corresponding density implies that 
Kepler-78b may have a structure that is deficient in iron and is thus
more like the Moon. Although the various approaches that were 
used to filter out the activity of Kepler 78 produce consistent radial velocity 
amplitudes to within the errors, these are still too large
to constrain the structure of this planet.
The uncertainty in the mass
for Kepler-78b is large enough to encompass models with structures ranging from
Mercury-like (iron enriched) to Moon-like (iron deficient). 
A more accurate $K$-amplitude as well as a better determination of the planet radius
are needed to distinguish between
these models.}


\titlerunning{The Mass of Kepler-78b}

   \maketitle
%

\section{Introduction}
Kepler-78b is a transiting Earth-sized planet in a 0.355-d orbit found by
the NASA Kepler Spacecraft (Sanchis-Ojeda et al.  2013). Although a careful analysis
of the exquisite Kepler light curves by Sanchis-Ojeda et al. (2013) established with high
probability that the transiting object was planetary in nature, final confirmation required
 Doppler measurements. This was established simultaneously by two teams.

Howard et al. (2013, hereafter H2013) took radial velocity (RV) measurements using the
HIRES spectrograph on the Keck telescope. They employed a version of harmonic
analysis where a series of sine functions were used to fit the RV signal due to activity. The
periods of the sine functions were restricted to the rotation period of the star, 
$P_{rot}$  and its harmonics
($P_{rot}$/2, $P_{rot}$/3, etc.). The  derived planet mass was 
1.69 $\pm$ 0.41 $M_\oplus$ (velocity $K$-amplitude = 1.66 $\pm$ 0.40
m\,s$^{-1}$).
This planet bulk density was 5.3$_{-1.6}^{2.0}$ g cm$^{-3}$ 
using a planet radius of $r_p$ = 1.20 $\pm$ 0.09 $R_\oplus$. 

Pepe et al. (2013, hereafter P2013) used RV measurements from the  HARPS-N
spectrograph to measure a $K$-amplitude of 1.96 $\pm$ 0.32 m\,s$^{-1}$, also based
on harmonic analysis of the rotation period.  This
resulted in a planet mass of $m_p$ = 1.86$_{-0.25}^{0.38}$  $M_\oplus$ and
a planet density of $\rho_m$ = 5.57$_{-1.31}^{3.02}$ g cm$^{-3}$ using
a planet radius of $r_p$ = 1.173$_{-0.089}^{0.159}$ $R_\oplus$. P2013 
also employed the
technique where the orbit was fit by allowing the
nightly velocity offset to vary (see below). This technique was first used to determine the mass
of the transiting planet, CoRoT-7b (Hatzes et al. 2010).  P2013 derived 
a velocity $K$-amplitude of 2.08 $\pm$ 0.32 m\,s$^{-1}$, a value consistent with
the harmonic analysis of the same data set and the value of H2013.
Both density determinations
firmly established Kepler-78b as a rocky planet.
The remarkable aspect of the RV detection of the planet was the fact that the host star, Kepler-78,
is modestly active star showing activity RV ``jitter'' of $\approx$ $\pm$ 10 m\,s$^{-1}$. The activity
signal thus dominates the RV reflex motion caused by the planet by a factor five or more. 

The RV detection of Kepler-78 b was made possible because
ultra-short period planets offer us a way to extract their signals from RV data that
are plagued by stellar activity noise.
If the orbital period of the planet is much smaller than the rotation period of the star, the nominal
timescale  of activity, it is possible to detect the planetary signal.
Kepler-78b has an orbital
period of mere  8.5 hours, much smaller than the stellar rotational period of 12.8 d (H2013).
Extracting short period planet signals from RV data of active stars was first demonstrated
for CoRoT-7b, a planet with a 0.85 d period orbiting an active star whose rotation 
period is 23 d (Queloz et al. 2009; Hatzes et al. 2010).

\begin{figure}
\resizebox{\hsize}{!}{\includegraphics{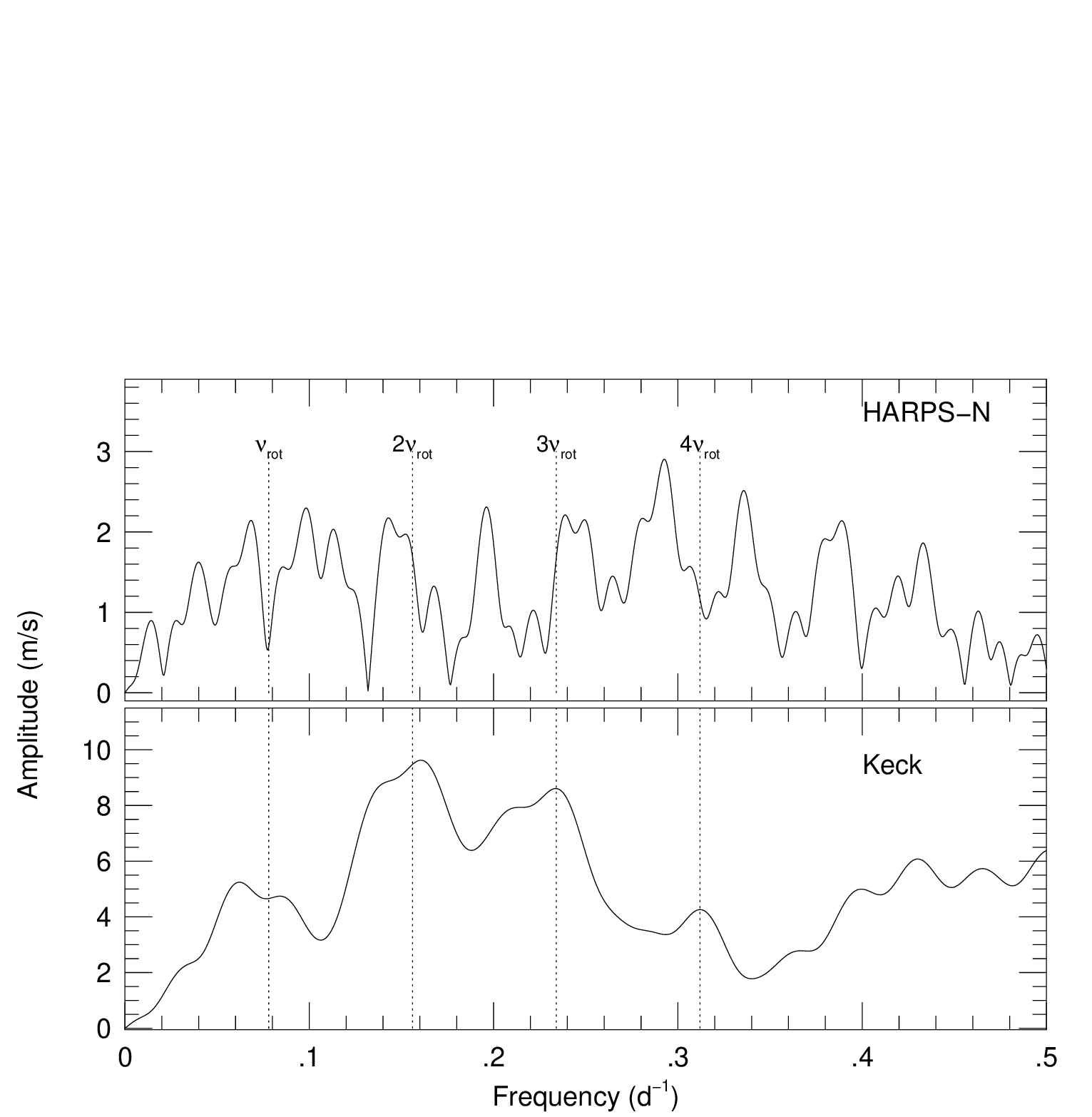}}
\caption{(Top) The low frequency Fourier amplitude spectrum of the
HARPS-N RV data. The dashed vertical lines mark the location
(left to right) of the rotational frequency, $\nu_{rot}$ and its first
three harmonics (2$\nu_{rot}$, 3$\nu_{rot}$, 4$\nu_{rot}$).
(Bottom) Same for the Keck RV data.
}
\label{dft1}
\end{figure}

In this paper we present an independent analysis of all the available
RV data for Kepler-78. Our goals are 
\begin{enumerate}
\item{To determine a refined mass using the combined HARPS-N and Keck radial velocity measurements}
\item{To assess the robustness of the detected planets to different approaches used in 
 filtering out the activity signal}
\item{To assess whether a Kepler-78b-like exoplanet can be detected in the RV data and accurate
orbital elements derived without the
transit information.}
\end{enumerate}

\section{Period Analysis}

\begin{figure}
\resizebox{\hsize}{!}{\includegraphics{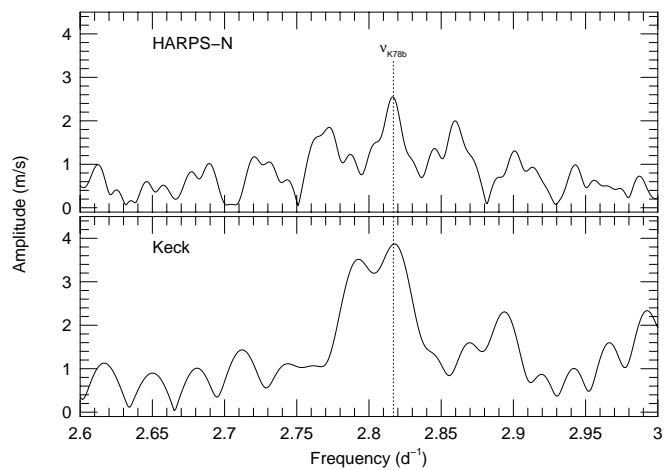}}
\caption{(Top) The Fourier amplitude spectrum of the
HARPS-N RV data centered on the orbital
frequency of Kepler-78b ($\nu_{K78b}$, dashed vertical line). 
(Bottom) The same for the Keck RV data. }
\label{dft2}
\end{figure}

\begin{table}
\begin{center}
\caption{HARPS-N Prewhitened Frequencies and Amplitudes} 
\begin{tabular}{cccc}
\hline
Period    &  Amplitude    & Phase & Comment  \\
(d)   & (m\,s$^{-1}$) &       &             \\
\hline
\hline
3.425  $\pm$   0.019  &   3.60 $\pm$ 0.39 & 0.837 $\pm$ 0.015 & $\approx P_{rot}/4$\\
10.405 $\pm$   0.089  &   4.26 $\pm$ 0.43 & 0.185 $\pm$ 0.013 & $\approx P_{rot}$ \\
0.3549 $\pm$   0.003  &   1.88 $\pm$ 0.44 & 0.729 $\pm$ 0.029 & $P_{K78b}$ \\ 
6.2578 $\pm$   0.480  &   1.62 $\pm$ 0.48 & 0.035 $\pm$ 0.034 & $P_{rot}/2$ \\
\hline
\end{tabular}
\end{center}
\tablefoot{$P_{rot}$ is the rotation period of the star
and $P_{K78}$ is the orbital period of Kepler-78b. The phase is reckoned with respect to the time of the
first HARPS-N measurement}
\end{table}

\begin{table}
\begin{center}
\caption{Keck Prewhitened Frequencies and Amplitudes}
\begin{tabular}{cccc}
\hline
Period    &  Amplitude    & Phase & Comment  \\
(d)   & (m\,s$^{-1}$) &       &             \\
\hline
\hline
6.443  $\pm$ 0.0370 & 11.92 $\pm$ 0.45 & 0.48 $\pm$ 0.013 & $P_{rot}/2$ \\
4.235  $\pm$ 0.0897 &  9.67 $\pm$ 0.62 & 0.12 $\pm$ 0.014 & $P_{rot}/3$\\
0.3549 $\pm$ 0.0024 &  2.01 $\pm$ 0.43 & 0.68 $\pm$ 0.037 & $P_{K78b}$ \\
11.312 $\pm$ 0.171  &  3.66 $\pm$ 0.41 & 0.31 $\pm$ 0.029 & $P_{rot}$ \\
\hline
\end{tabular}
\end{center}
\tablefoot{The phase is reckoned with respect to the time of the
first Keck measurement.}
\end{table}

\subsection{Prewhitening}

Fourier component analysis via the so-called prewhitening procedure is a standard way of extracting
multi-periodic signals from time series data. It selectively identifies the Fourier components of a
time series by first finding the dominant frequency, fitting a sine wave to the data and removing this,
and then searching for the next highest Fourier component. An obvious application 
is the detection of multi-periodic planet systems (e.g. Hatzes 2013a), but it can also be
an effective tool for filtering out activity noise that  may not be strictly periodic (Hatzes et al. 2010, 2013b).
{We should note that harmonic analysis based on the
stellar rotation frequency  is a form of prewhitening. 
However, it restricts the  choice of 
frequencies only to that of rotational frequency and its harmonics}. 
With traditional prewhitening, on
the other hand,  one can select the 
highest amplitude frequency found in the time series or its residuals, 
even if it is not obviously related to the rotation of the star.

We applied prewhitening individually to the HARPS-N and Keck data sets in order
to evaluate the Fourier spectrum of each time series and to test how effective this procedure is in
extracting  the planetary signal. All data values were used, including nights where only one RV measurement
was made. 

Figure~\ref{dft1} shows the low frequency Fourier amplitude spectrum of each data set 
centered around the expected stellar rotational frequency. Vertical dashed lines mark the
location of the rotational frequency, $\nu_{rot}$, as well as its first 3 harmonics
( 2$\nu_{rot}$,  3$\nu_{rot}$, and  4$\nu_{rot}$).\footnote{The convention for this paper is to start the numbering
of the harmonics with the first ``overtone''  2$\nu_{rot}$ ($P_{rot}$/2), rather
than with the rotational frequency.} 
Note that the Keck RV data are clearly dominated
by the first harmonic, but it also shows strong evidence for  the presence
of 3$\nu_{rot}$.
The rotational frequency and 4$\nu_{rot}$ are also present, but only weakly so.
The Fourier amplitude spectrum lends support to the use of harmonic analysis on the Keck data.
H2013 only used components out to 3$\nu_{rot}$ since including higher terms did not improve the 
solution. This is also supported by the Fourier spectrum as the amplitude of 4$\nu_{rot}$ significantly drops
from that of 3$\nu_{rot}$.

The HARPS-N data, on the other hand, show no clear evidence for the rotational frequency
or any of the harmonics. The dominant frequency at $\nu$ = 0.29 d$^{-1}$  might
be related to 4$\nu_{rot}$, which implies a rotational period of 13.8 d, much higher than the rotational 
period found by H2013.

Tables 1 and 2 list the frequencies found by the prewhitening process for the HARPS-N
and Keck data, respectively. Both produce comparable
rms scatter about the fit: 2.57 m\,s$^{-1}$ for HARPS-N and 
2.51 m\,s$^{-1}$ for Keck. We can exploit the fact that the Keck data  clearly
show evidence for the presence of rotational harmonics to determine the rotational period.
If the ``non-planetary'' frequencies are indeed due to rotational harmonics
then the observed low frequencies should occur at intervals
of $\nu_{obs}$ $ =$ $n$$\nu_{rot}$, where $\nu_{rot}$ 
is the rotational frequency and $n$ an integer. The value  of $\nu_{rot}$ that
minimized $\sum_{n=1}^3 (\nu_{obs} - n\nu_{rot})^2$  corresponded to 
$P_{rot}$ = 12.7 $\pm$ 0.3 d, consistent with the value of H2013.

What is surprising is that pre-whitening of the HARPS-N data
produced essentially the same $K$-amplitude even
though standard prewhitening found differerent frequencies. The 10.4 d period is
near the $\sim$ 10 d period used by P2013, whereas the 3.42 d period is close to,
but not coincident with the 4.2 d period used by P2013. The 6.2 d period
found by pre-whitening was not mentioned by P2013. Regardless of the fact
that different Fourier components for the activity signal have been removed,
the same amplitude is derived: 1.88 $\pm$ 0.44 m\,s$^{-1}$ for prewhitening
compared to 1.96 $\pm$ 0.32 m\,s$^{-1}$ for harmonic analysis. 
This speaks for the robustness
of the Kepler-78b detection, an issue that we shall return to later. 

An orbital solution using the transit ephemeris and zero
eccentricity was made to the combined residual data after removing the activity
signal (i.e. frequencies) from the data. The solution resulted
in a K-amplitude of 1.90 $\pm$ 0.23 m\,s$^{-1}$ which is consistent with both
P2013 and H2013.

\begin{figure}
\resizebox{\hsize}{!}{\includegraphics{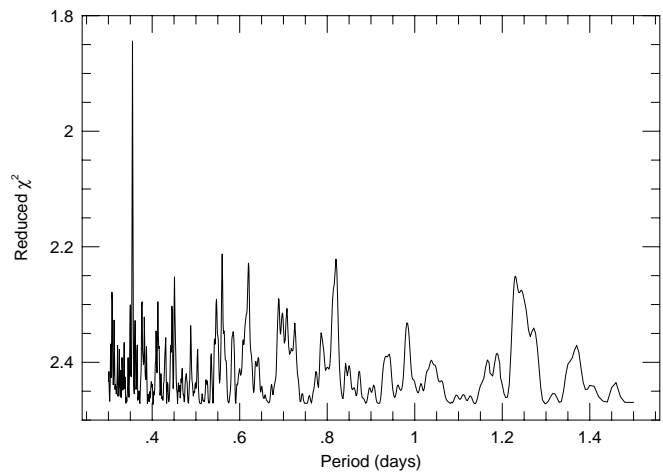}}
\caption{The FCO periodogram of the combined Kepler-78 RVs (HARPS-N plus
Keck). The scale of the y-axis has been reversed so that  a lower
$\chi^2$ appears as a peak.
}
\label{fcop_k78}
\end{figure}

\subsection{The Floating Chunk Offset Periodogram}

The RV detection of Kepler-78b was abetted by the fact that a periodic
transit signal had been found in the photometry that was validated
as due to an orbiting planet (Sanchis-Ojeda et al. 2013). 
We thus have the luxury of knowing both the
orbital period and orbital phase -- we only need to fit 
the  velocity amplitude.
The question naturally arises, ``Could Kepler-78b had detected without
a priori knowledge of a transiting planet?"

There is already evidence in the raw Fourier amplitude spectrum for the presence of Kepler-78b.
These show what appears to be a significant peak precisely at the orbital
frequency of Kepler-78b (Fig~\ref{dft2}). In both the HARPS-N and
Keck data this peak is approximately four times the height of the surrounding noise floor.
Kuschnig et al. (1997) established through simulations that this corresponds
to a false alarm probability of about 1\%. The planet signal is reasonably significant
without prior knowledge of the transit period.  However, in the case of
active stars the Fourier spectrum can be completely dominated by the numerous
frequencies arising from activity that make
it difficult to assess the true statistical significance.

Hatzes et al. (2010) introduced a method by which one could extract the signal
of short period planets even in the presence of considerable activity noise. The
method exploits cases where the planet orbital period is significantly less
than periods arising from activity. Depending on the rotational period of the star,
during one night the RV contribution by surface structure (spots, plage, etc.) is
small since the star has had little time to rotate, or even for spots to evolve.
The spot configuration on the star is essentially frozen-in and it only adds a constant
value to the observed RV. When taking a long time series of measurements
throughout the night one should only see changes in the RV due to the orbital motion
of the planet. Fitting these RV measurements with a Keplerian orbit, but allowing
the zero points of the nightly data to  vary is an effective way
to filter out the activity variations and to derive an accurate measurement of the
K-amplitude. We shall refer to this technique as the Floating Chunk Offset (FCO) method.

FCO was successfully applied to the first transiting rocky planet CoRoT-7b,  a case where the 
K-amplitude of the planet was 2-4 times lower than the variations due to activity
(Hatzes et al. 2010, 2011). In the case of Kepler-78b we are hindered by the fact
that the K-amplitude is a factor of 5-10 times lower than the activity signal,
but we are helped by a much shorter orbital period  for Kepler-78b (0.355 d compared to 0.85 d
for CoRoT-7b).

{FCO can also be used as a form of a ``periodogram'' (Hatzes et al. 2010, 2011)
to search for planetary signals in the presence of activity noise. 
For a transiting planet the orbital period is known, so a periodogram is of limited
use, except possibly for investigating the signficance of the signal.
The true utility of a periodogram, however, is finding unknown planets in your
time series. For these we do not know the phase of the orbit (e.g. transit time), the
period, or the eccentricity.

To explore the behavior of the FCO as a ``periodogram'' we fitted the data using a trial 
orbital period 
and allowed the nightly
offsets to vary. The trial period was then varied over a range of values and the
goodness-of-fit examined. The eccentricity
was fixed to zero since circular orbits are a good assumption for short period planets.
However, it is possible to vary the eccentricity once a best-fit period is found.
We also allowed the orbital phase to vary because another period with a different
phase might provide a better fit to the data than using the transit phase.
For each solution we computed the reduced $\chi^2$  and then
repeated this for another trial period. The lowest value of $\chi^2$ in the 
$\chi^2$ -- Period diagram  
revealed the strong  periodic signal present in the data. The FCO fit 
was performed using the least squares fitting program
{\it Gaussfit} (Jefferys et al. 1988). 
}

Figure~\ref{fcop_k78} shows this FCO  periodogram 
as applied to combined Kepler-78b RV data. Note that even though
the Keck and HARPS-N have different zero point offsets, this is irrelevant 
to FCO as it calculates velocity offsets for separate nights. It thus offers a natural
way of combining data taken with different instruments and velocity offsets
in a least squares sense. 
In this
figure the scale of the y-axis is reversed so as to mimic a classic periodogram
where a peak signifies high power, and thus a signifcant signal. In our case  the peak
of the FCO periodogram
is not a power, but goodness of fit.
We plot $\chi^2$ as opposed to 1/$\chi^2$ so that the reader can get
an immediate sense for the quality of the fit.
The $\chi^2$ shows a clear peak (minimum) at the orbital period of
Kepler-78b.

\subsection{Tests of the FCO Periodogram}

\begin{figure}
\resizebox{\hsize}{!}{\includegraphics{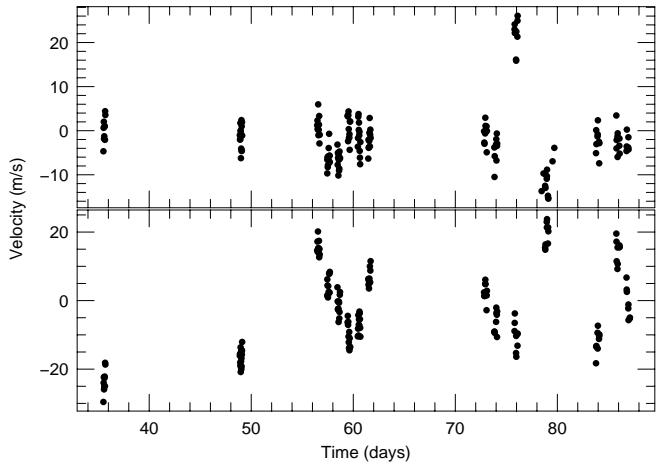}}
\caption{(Top) Simulated data consisiting of a simulated activity
signal (Model 1), a planet signal, plus random noise with
$\sigma$ = 2.4  m\,s$^{-1}$. (Bottom) Simulated data using
Model 2 for the activity signal. }
\label{activity}
\end{figure}

The FCO method was tested on simulated activity signals to see how
well it could recover a known planet signal inserted into the data. To do
this we added the Kepler-78b orbit to simulated RV data sets generated by two
activity ``models''. The first (Model 1) consisted of 
an activity signal generated using the frequencies found in the
prewhitening process. The respective prewhitened frequencies for
each data set were used (Table 1 for the ``HARPS-N'' data and Table 2 
for the ``Keck'' data), but without the planet signal. The orbital solution
was then inserted into the activity signal with $K$ = 1.5 m\,s$^{-1}$. These
simulated data were sampled the same way as the real data and
random noise at a level of $\sigma$ = 2.4 m\,s$^{-1}$ was also added.
The top panel of Figure~\ref{activity} shows the simulated data
set.

\begin{figure}
\resizebox{\hsize}{!}{\includegraphics{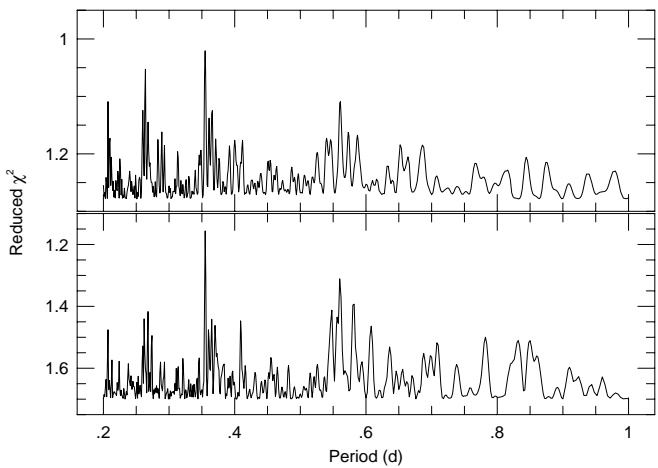}}
\caption{(Top) FCO periodogram of the simulated
data set from Model 1 for the activity plus planet. (Bottom)
The FCO periodogram for  Model 2 plus planet.
}
\label{fcop}
\end{figure}

\begin{figure}
\resizebox{\hsize}{!}{\includegraphics{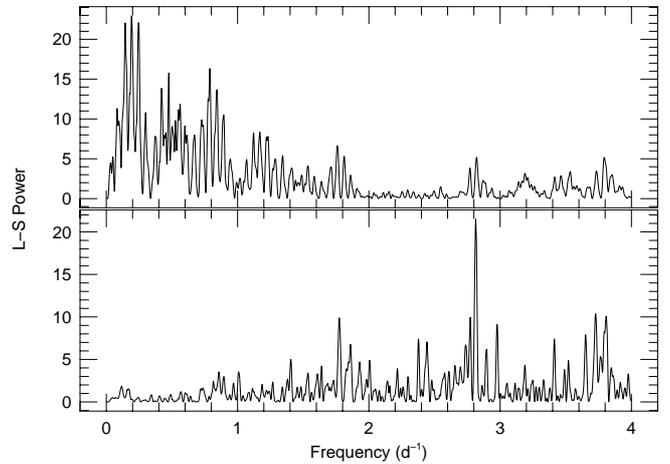}}
\caption{(Top) The L-S periodogram of the simulated
Model 1 data. (Bottom) The L-S perodogram of simulated
Model 1 data after removing the calculated nightly offsets.
}
\label{a1scargle}
\end{figure}

The resulting floating chunk offset periodogram is shown in the top
panel of Figure~\ref{fcop}. Again, in this case
we also allowed the orbital phase for each trial period to vary.
The highest peak (minimum $\chi^2$) corresponds
to the planet frequency. The top panel of Figure~\ref{a1scargle} shows
the Lomb-Scargle (L-S) periodogram (Lomb 1976; Scargle 1982) of the original data, 
We use the L-S periodogram
in its original form rather than the generalized L-S periodogram (GLS, Zechmeister \& K\"urster 2009)
because we prefer to show the unnormalized power -- the larger
the power, the higher the significance of a peak. For data sets with a large
number of points the L-S periodogram  produces consistent results as GLS.
Note that one clearly see the signal of the planet in the  unfiltered L-S periodogram
in spite of the activity signal. The lower panel shows
the L-S periodogram after applying the nightly offsets from the best
fit period. The offset fitting has acted as a high pass filter that has suppressed
the low frequency components due to activity and greatly increased the power
at the planet orbital frequency. The recovered $K$-amplitude was $K$ = 1.29 $\pm$ 0.26
m\,s$^{-1}$.

\begin{table}
\begin{center}
\caption{$K$-amplitudes}
\begin{tabular}{cccccl}
\hline
N  &   K                       &    O-C$_{H}$ & O-C$_{K}$ & $\chi^2$ &  Comment \\
     & (m\,s$^{-1}$)      &  (m\,s$^{-1}$) & (m\,s$^{-1}$) &           & \\
     \hline
1   & 1.96 $\pm$ 0.32 & 2.34       & $-$                  &  -  & P2013 \\
2   & 1.66 $\pm$ 0.40 & $-$         & $-$                  & 1.12  & H2013 \\
3   & 1.90 $\pm$ 0.32 & 2.59        &  2.55                 & 2.60  & Prewhitened \\
4   & 1.86 $\pm$ 0.30 & 2.11        &  2.55                 & 1.03  & H-N \\
5   & 1.72 $\pm$ 0.40  & $-$         &  2.58                 & 2.49  & Keck \\
6   & 1.77 $\pm$ 0.26  & $-$         &  2.58                 & 2.49  & All, no Jitter \\
7   & 1.63 $\pm$ 0.23  & 2.12         & 2.58                 & 1.02   & All w/Jitter \\
8   & 1.30 $\pm$ 0.35  & $-$         & 2.58                   & 1.00 &  Keck w/Jitter \\
9   & 1.39 $\pm$ 0.37  & $-$          & 2.08                  & 0.95  &  H-N, $N$ $>2$ \\
10   & 1.34 $\pm$ 0.25  & 2.08        & 2.59                   & 0.95  & Adopted \\       
\hline
\label{solutions}
\end{tabular}
\end{center}
\label{modtab}
\end{table}

\begin{table}
\begin{center}
\caption{Jitter Values for the Keck data}
\begin{tabular}{cccc}
\hline
Dates                              &  Mean Error   & O-C & Jitter  \\
         (JD - 24400000)  & (m\,s$^{-1}$) & (m\,s$^{-1}$)  & (m\,s$^{-1}$)   \\
\hline
\hline
48.88 - 49.11	  & 1.95  & 2.87   & 2.10 \\
72.82 - 73.10	  & 1.89  & 2.63   & 1.83 \\
73.83 - 74.10	  & 1.69  & 3.22   & 2.74 \\
75.80 - 66.11	  & 1.64  & 1.68   & 0.00 \\
78.80 - 79.12	  & 1.60  & 2.43   & 1.83 \\
83.79 - 84.12	  & 1.74  & 3.12   & 2.59 \\
85.78 - 86.10	  & 1.64  & 3.15   & 2.69 \\
86.78 - 87.11	  & 1.74  & 1.92   & 0.81 \\
\hline
\end{tabular}
\end{center}
\label{jitter}
\end{table}

For the second model of the activity variations we considered
a more complicated case. The times of the data were divided
into three epochs: JD-2440000 = 35 -- 49, 56 -- 61, and 72 -- 87.
In each epoch we generated a simulated activity 
signal using a sum of sine functions but with slightly different rotation periods and harmonics in
each epoch.
The periods and amplitudes for the sine functions were as follows.
Epoch 1: $P_1$ = 12.83 d, $K_1$ = 20.7 m\,s$^{-1}$; $P_2$ = 6.42 d, 
$K_2$  = 10.6 m\,s$^{-1}$. Epoch 2:
$P_1$ = 11.50 d, $K_1$ = 9.0 m\,s$^{-1}$; $P_2$ = 6.42 d, $K_2$ = 15.0 m\,s$^{-1}$; 
$P_3$ = 3.70 d, $K_3$ = 4.6 m\,s$^{-1}$. Epoch 3: $P_1$ = 13.61 d, $K_1$ = 5.0 m\,s$^{-1}$;
$P_2$ = 6.8 d, $K_2$ = 10.0 m\,s$^{-1}$; $P_3$ = 3.44 d, $K_3$ = 8.6 m\,s$^{-1}$.
Note that in in each epoch slightly different ``rotational'' periods and harmonics
were used.  This was
to mimic spots located and emerging at different latititudes of a differentially
rotating star. We should point out this might present difficulties to the harmonic analysis which
assumes a single rotation period and its harmonics.

The lower panel of Figure~\ref{activity} shows the resulting  simulated
activity signal. This includes the planet signal as well as random noise
with $\sigma$ = 2.4 m\,s$^{-1}$. Notice that this model produces much
larger variations for the activity signal, $\Delta$RV = $\pm$ 20 m\,s$^{-1}$.
The resulting FCO periodogram is shown in the lower panel of Figure~\ref{fcop}.
Again, the $\chi^2$ is minimized (appearing as the highest peak) for the correct
orbital period.  The fitted $K$-amplitude was $K$ = 1.48 $\pm$ 0.27
m\,s$^{-1}$.

\begin{figure}
\resizebox{\hsize}{!}{\includegraphics{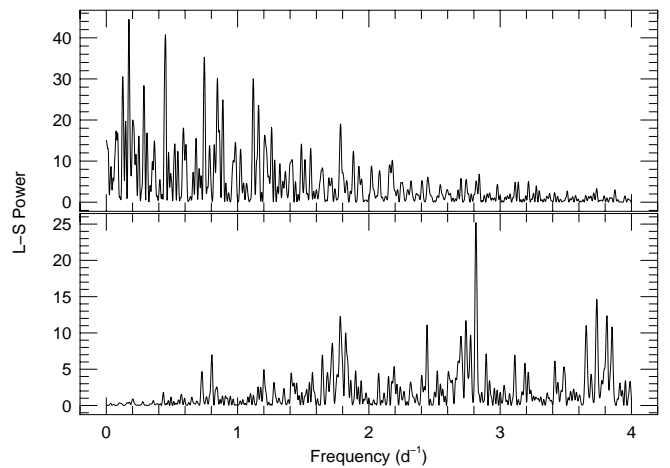}}
\caption{(Top) The L-S periodogram of the simulated
Model 2 data. (Bottom) The L-S perodogram of simulated
Model 2 data after removing the calculated nightly offsets.}
\label{a2scargle}
\end{figure}

Figure~\ref{a2scargle} shows the L-S periodogram
of the original (top) and filtered (bottom) RV data. Note that there is virtually no power
at the orbital frequency of 2.817 d$^{-1}$ in the original data, but that this becomes
very prominent in the power spectrum of the filtered data. Again, the numerous
peaks at low frequency due to activity are effectively  suppressed. 

\begin{figure}
\resizebox{\hsize}{!}{\includegraphics{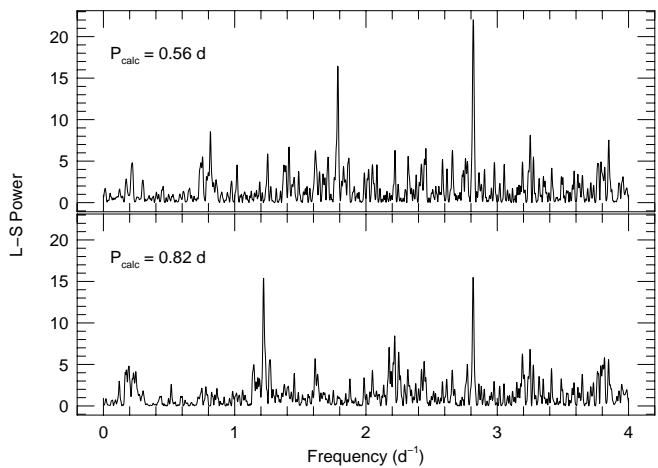}}
\caption{The L-S periodogram after using the FCO method on the 
RV data with a period of 0.56 d (top) and 0.82 d (bottom). Note that the  orbital frequency of Kepler-78b
($\nu$ = 2.817 d $^{-1}$)
is still the dominant peak}
\label{robust}
\end{figure}

As a final check of the robustness of the planet signal we took the
Keck + HARPS RV data and applied the chunk offset fitting method, but used the {\it wrong}
orbital period, {i.e.  a period
other than 0.355-d,  in calculating the offsets.} Two trial 
periods were used in the fitting,  the second and third minima
(peaks) in the FCO periodogram, namely $P$ = 0.56 d and 0.82 d.
Figure~\ref{robust} shows the  L-S periodogram of the
RV data after removing the nightly offsets calculated with the incorrect orbital periods.
Remarkably, the true orbital frequency at 2.817 d$^{-1}$ still comes through
as the dominant peak. In the case of the 0.56 d fit, it is clearly the dominant peak.
In the 0.82-d fit, it is comparable to the spurious peak at the frequency
corresponding to the period (0.82-d) used in calculating the nightly offsets.
This speaks for the robustness of the planet signal found by the FCO method.

\begin{figure}
\resizebox{\hsize}{!}{\includegraphics{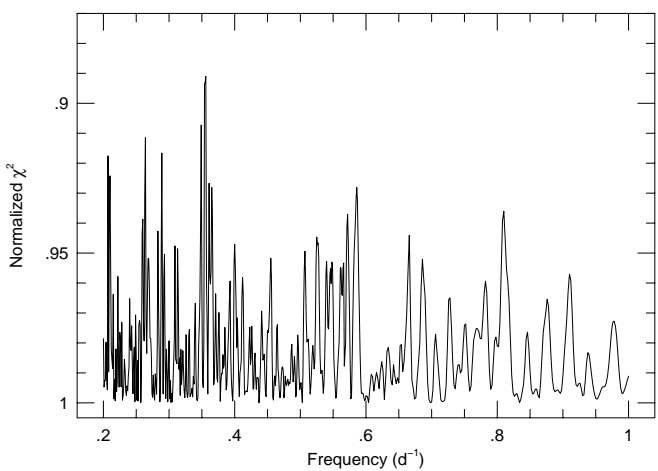}}
\caption{The FCO periodogram using the second activity model and the planet orbit added with 
$K$ = 1.0 m\,s$^{-1}$. The highest peak coincides to the orbital frequency of 
$\nu$ = 2.817 c\,d$^{-1}$.}
\label{k1per}
\end{figure}

To explore the detection limit of a Kepler-78b-like planet in the RV data, we took the first
activity model (Model 1) as it best represents the variations seen in the actual data. We 
then added the planet signal with $K$ = 1.0 m\,s$^{-1}$ and noise at a level of 
2.4 m\,s$^{-1}$. Figure~\ref{k1per} shows the resulting FCO periodogram.
It is considerably noiser, but the dominant peak still occurs at the planet frequency.

\begin{figure}
\resizebox{\hsize}{!}{\includegraphics{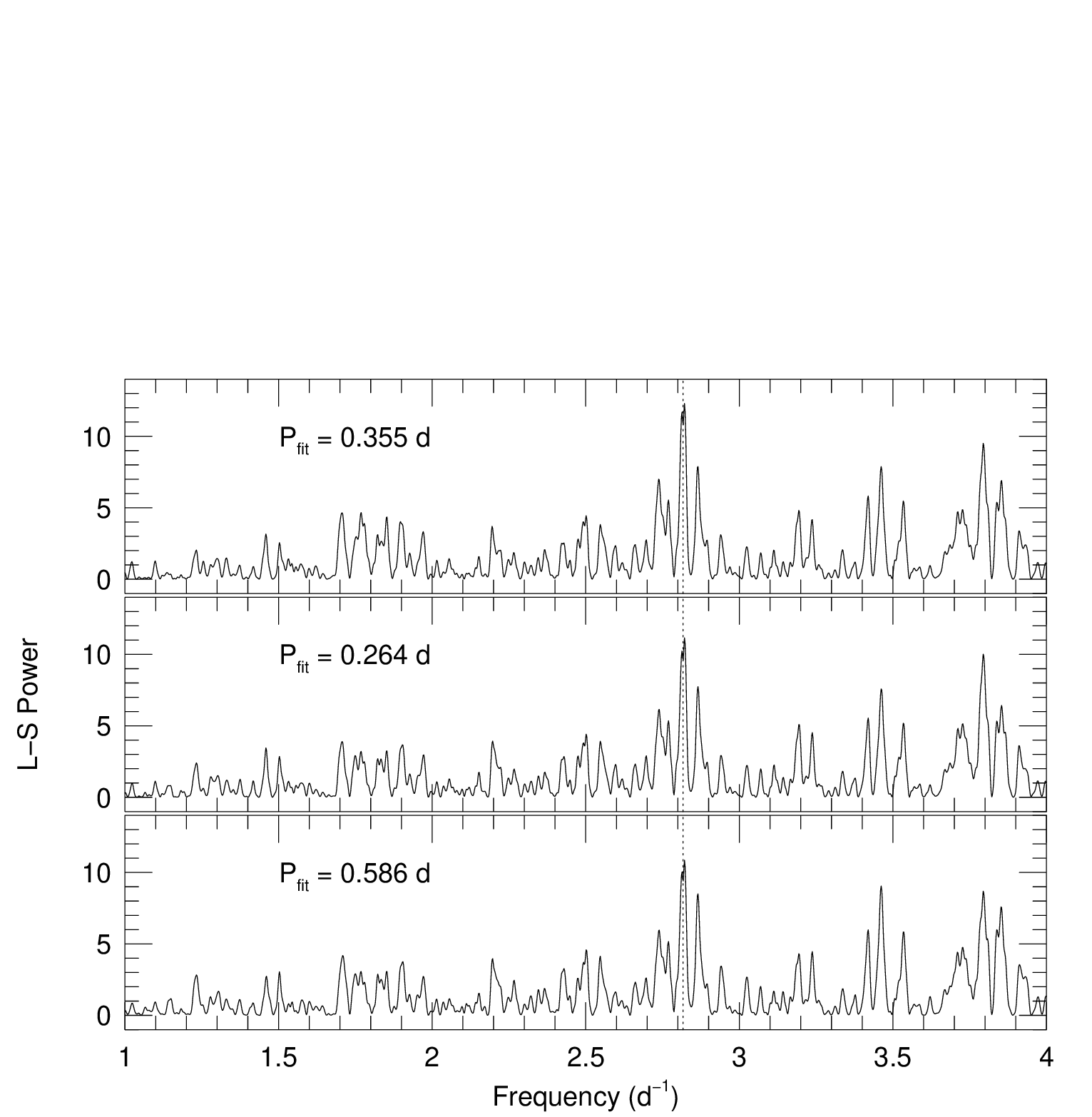}}
\caption{The L-S periodogram of the simulated data with a planet
$K$-amplitude of 1.0 m\,s$^{-1}$ and after subtracting the calculated
offsets. (top) Offsets calculated using the true planet orbital period. The lower panels
show the results after using the wrong orbital periods for calculating the offsets,
0.264 d (middle) and 0.586 d (bottom). The dominant peak is still 
at the true orbital frequency. }
\label{k1ft}
\end{figure}

The top panel of Figure~\ref{k1ft} shows the L-S periodogram after fitting
these synthetic data with the 0.355-d period and subtracting the computed offsets. The dominant
peak is at the correct orbital frequency for the planet. We also show
two other periodograms where in these cases we calculated  the nightly  offsets using the
wrong ``orbital'' periods of
0.264 d and 0.586 d. These periods coincide with local minima (i.e. peaks in the periodogram).
The dominant peak remains at the true value of
$\nu$ = 2.817 d $^{-1}$.

\section{The Mass of Kepler-78b}

An orbital fit was made to the combined RV data using the FCO
 technique in order to get a better determination
of the $K$-amplitude.
Three approaches were made to fitting the data:
1) the data in its original form and the estimated errors, 2) trend fitting on
nightly data with the addition of  a so-called jitter correction, and 3) fitting only those
nights for which multiple measurements ($N$ $>$ 2, but typically
6--10) were made. 

First, an FCO fit was made to the individual data sets keeping the
ephemeris (epoch of transit and orbital phase) fixed to the values
derived from the the transit light curve. The HARPS-N (H-N) only
data resulted in $K$ = 1.86 $\pm$ 0.30 m\,s$^{-1}$,
observed minus computed ($O-C$) = 2.11 m\,s$^{-1}$,
and a reduced $\chi^2$ = 1.03 (Solution 4 in Table 3). These values are fully consistent
with those published by P2013.  

{We also investigated allowing the orbital phase to be a free parameter. If the derived
orbital phase is consistent with the transit values this would provide even stronger
confirmation that the RV variations are indeed due to a planet. 
For circular orbits the angle of periastron, $\omega$, is ill-defined, but it can
still be used as a measure of the orbital phase. 
For eclising systems where the epoch is defined as the primary eclipse (mid-transit time in our case)
it should be exactly 90$^\circ$  (see Phoebe 2.0.0 documentation, www.phoebe-project.org).
So, when fitting the orbit to a fixed epoch, the phase of the orbit is represented by the
parameter $\omega$. 
In this case, a value of $\omega$ significantly different from 90$^\circ$ would indicate
that the RV data is better fit with a circular orbit not in phase with the time of mid-transit
and this would cast some doubt on the reality of the RV signal.

Allowing the orbital phase to vary resulted 
in $\omega$ =  88.6 $\pm$ 11.9  degrees, a value consistent
with the transit ephemeris and strong confirmation that the RV variations are due
to the planet. Had one discovered Kepler-78b with RV measurements first,
the HARPS-N data are of sufficient quality
to predict the 
time of transit to within 0.035 orbital phase or 18 minutes.

The fit using only the Keck data and again fixing  the orbital phase
to the transit value resulted in $K$ = 1.72 $\pm$ 0.40 m\,s$^{-1}$,
$O-C$ = 2.58 m\,s$^{-1}$, and $\chi^2$ = 2.49 (Solution 5 in Table 3). These values are
consistent with those of H2013. Allowing the orbital phase to be  a free parameter resulted in
$\omega$ = 74.6 $\pm$ 16.7 degrees, again in agreement
with the transit ephemeris to 1$\sigma$. Given the short
orbital period the Keck solution could predict the transit time to within 20 minutes.}

Finally, fitting all the Keck  + Harps-N radial velocity data yielded  
$K$ = 1.77 $\pm$ 0.26 m\,s$^{-1}$ (Solution 6).

H2013 noted that on some nights the Keck data had
linear trends that were removed before computing the orbit. We performed
a visual inspection to the data and noticed that only on Night
4 (JD $-$ 2456400 $=$ 75.8 -- 76.10) 
was there a significant  linear trend in  the data. 
The other nights
showed only marginal trends that were difficult to discern
due to the scatter of the measurements. This was confirmed
by performing a linear least squares fit to the nightly
data. On Night 4 the slope in the RV variations was
more than four times larger than its error.
On all other nights the slope in the RV variations had a value of
zero to within the errors. 
Therefore, for the other nights we chose {\it not} to
detrend the data as these might adversely affect the $K$-amplitude. The Keck data 
typically cover around 0.3 of an orbital phase and during this time
the orbital motion itself might appear as a linear trend. The same, of course,
holds for Night 4, however the trend in this case was large ($\Delta$RV = 5 m\,s$^{-1}$)
and all measurements during the night followed the linear trend.
We can thus be reasonably confident
of fitting mostly the variations of the trend and not
suppressing the variations due to the orbital motion. 

H2013 also increased the measurement error by 
applying  a ``jitter'' correction to account for the actual errors being 
larger than the estimated ones. The fact that the  reduced $\chi^2$  
of the Keck solution is much larger than unity (Table 3) indicates
that there may be additional sources of error in the Keck measurements.
H2013 applied an overall jitter correction of 2.1 m\,s$^{-1}$
that was added in quadrature to the nominal errors. The source
of this jitter is not clear, but it most likely due to an underestimate
of the errors, or the presence of unknown systematic errors. It is unlikely
that this is due to intrinsic stellar noise as the value of near unity for the
reduced $\chi^2$ of the
 HARPS-N radial velocities imply that the actual errors are consistent
 with the estimated ones.

Regardless of whether the additional jitter results from systematic errors
or stellar intrinsic variations there is no reason to expect that this jitter
value should be the same on  each night. Rather than applying a single overall jitter correction, we applied
a {\it nightly} jitter correction term determined in a bootstrap fashion. First, an orbital solution was made
using the combined data and nominal errors. The rms scatter of the
residuals on a given night was then compared to the mean error.
If the actual rms was less than or equal to the mean error, no jitter
was added.  Otherwise, a   jitter correction was added
in quadrature with the estimated error  in order to make the rms scatter about
the orbital fit consistent with the jitter-corrected errors for that night 
(Table 4).
Note that the mean value is consistent with the value of 2.1 m\,s$^{-1}$ used
by H2013.
No jitter correction was applied to the  HARPS-N data as the reduced
$\chi^2$ for that data set was already close to unity.

The fit to the combined RV data resulted in $K$ = 1.63 $\pm$ 0.23 m\,s$^{-1}$
and a corresponding planet mass of 1.59 $\pm$ 0.22 $M_\oplus$. This is Solution 7 in Table 3.
The application of the jitter values to the Keck data produced a $\chi^2$ of nearly
unity. This $K$-amplitude is consistent with both the H2013 and P2013 values with
the errors, but our nominal value is closer to the H2013 value.

\begin{figure}
\resizebox{\hsize}{!}{\includegraphics{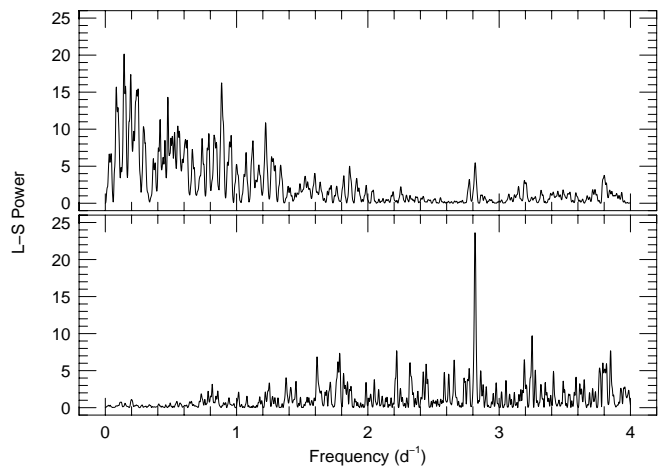}}
\caption{(Top) The L-S periodgram of the combined
Kepler-78 RVs (HARPS-N plus Keck). An offset has been
applied to the HARPS-N data so that both data sets have the
the same zero-point. (Bottom) The L-S periodogram
of the combined data sets after applying the calculated
nightly offsets.
}
\label{K78scargle}
\end{figure}

The lower panel of Figure~\ref{K78scargle} shows the L-S periodogram
of the RVs after removing the nightly offsets. 
The dominant peak occurs exactly at the orbital frequency of $\nu$ = 
2.817 d$^{-1}$. The signal appears to be significant with power, $z$ = 20
which implies a small false alarm probability (FAP $<$ 10$^{-6}$). However,
we caution the reader that the FAP may be artificially low since the data has been filtered.
In such cases a formal low FAP may not necessarily indicate a high
statistical significance (Hatzes 2013).

The FCO technique works best when there are multiple measurements
that sample well the segment of the orbit
seen on a given  night.
The Keck RV data, although having a slightly poorer precision,
has the advantage over the HARPS-N data of better sampling with 9 to 16 measurements
taken each night. We noticed that the fit to the Keck data alone produced a slightly lower
$K$-amplitude of 1.30 $\pm$ 0.35 m\,s$^{-1}$ (Solution 8 in Table 3). This may be due to the fact
that the HARPS-N data had fourteen nights  on which only two RV measurements
were taken. This reflects the strategy by P2013 of taking many measurements
at the extrema of the RV curve where one has more ``leverage'' in terms of
detecting a change in the velocity. However, there are two drawbacks to this
strategy.  First, the  RV measurement error is comparable,
if not smaller than the $K$-amplitude, so two measurements may be insufficient
for getting a good RV measurement on a given night.
Second, only two measurements in a night provide poor sampling of the shape of the orbital
variations. Without the shape information part of the activity
related velocity offset may be absorbed into the RV amplitude and
could result in  a systematicaly higher or lower RV amplitude.

\begin{figure}
\resizebox{\hsize}{!}{\includegraphics{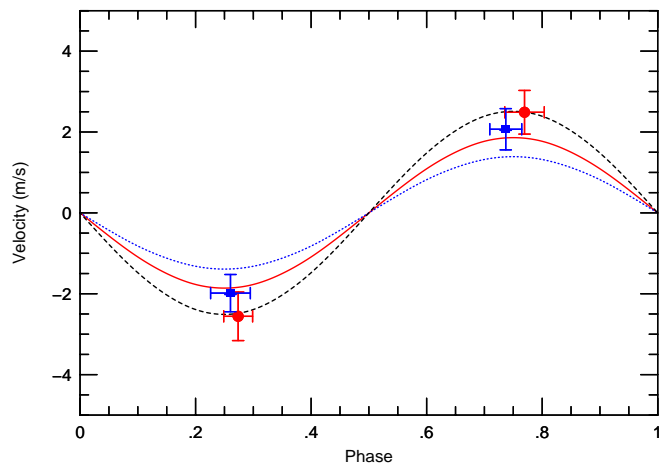}}
\caption{The phase-binned averages of the orbital variations of
Kepler-78b at phases 0.25 and 0.75, the extrema of the orbital
motion. The circles represents values using data from nights were
only two RV measurements were made. The squares represent
binned averages from nights where typically 6-10 measurements
were made. {The solid curve represents the orbital solution to the full
HARPS-N data set. The dotted line represents the fit using  a subset of the
HARPS-N measurements where more than two RV measurements on a night were taken.
The long-dashed line represents the best fit sine curve with amplitude $K$ = 2.5 m\,s$^{-1}$
that passes through the RV extrema represented by the circles.}
\label{2phase}}
\end{figure}

To check this we calculated
phase-binned averages around orbital phases 0.25 and 0.75. We treated the 
nights with more than two measurements separate to those
where only two measurements were made. The binned values from nights
with only two measurements show a slightly larger amplitude 
than that from 
nights with multiple measurements (Figure~\ref{2phase}). This is also confirmed
by calculating an orbit using only the HARPS-N data with more than two measurements
per night. The $K$-amplitude is smaller ($K$ = 1.39 $\pm$ 0.37 m\,s$^{-1}$, Solution 9) and is
consistent with the amplitude derived using only the Keck data.

{Note that for the HARPS RVs of Kepler-78 the best fitting sine curve that passes
through the phase-binned averages at phase 0.25 and 0.75 results in a rather
high $K$ amplitude of 2.5  m\,s$^{-1}$. In this case using only RV data taken at the extrema
of the RV curve can result in an estimate of the velocity amplitude that  can differ
by a factor of two from the value obtained by fitting the full RV curve.}

\begin{figure}
\resizebox{\hsize}{!}{\includegraphics{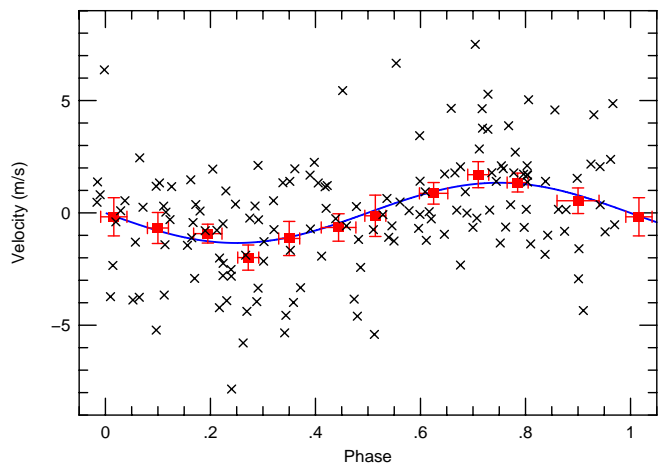}}
\caption{The phased offset-corrected RV measurements
of Kepler-78 RV data for which more than two
RV measurements per night were obtained (dots). 
The bin-averaged values are shown
as squares. The orbital solution is the curve.
The ``error'' on the phase is just the standard deviation of
the phases  used for the binned value.
}
\label{orbit}
\end{figure}

For our ``adopted'' solution we therefore used only RV data from those nights
where more than two RV measurements were taken. This resulted in a 
$K$-amplitude of 1.34 $\pm$ 0.25 m\,s$^{-1}$, a corresponding planet mass
of $m_p$ = 1.31 $\pm$ 0.24 M$_\oplus$ (Solution 10), and a density
of $\rho$ = $4.5_{-2.0}^{+2.2}$ g\,cm$^{3}$ using the planet radius from P2013. 
The RV values after subtracting the
calculated nightly offsets and phased to the orbital period are  shown
in Figure~\ref{orbit}. Also shown are phase-binned averages so that one
can see better the orbital motion.

\begin{figure}
\resizebox{\hsize}{!}{\includegraphics{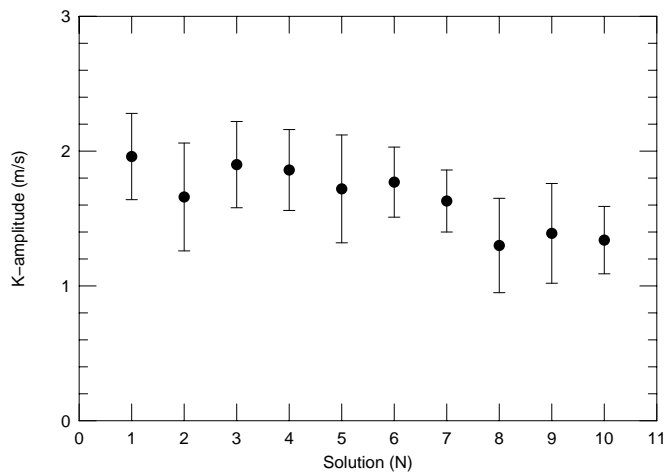}}
\caption{The various values of the $K$-amplitude using different
approaches to removing the activity. The "Solution'' number refers are listed in Table~\ref{modtab}
}
\label{models}
\end{figure}

All the $K$-amplitude determinations using the various methods including
the published values  and those from pre-whitening are
summarized in Table 3. For all our solutions the period and phase
were fixed to the transit values. A graphical version of the table is shown in Figure~\ref{models} where
the solution ``number'' refers to the first entry in the table. We note that
all determinations of the RV amplitude are consistent with each other
to within 1$\sigma$.

\begin{figure}
\resizebox{\hsize}{!}{\includegraphics{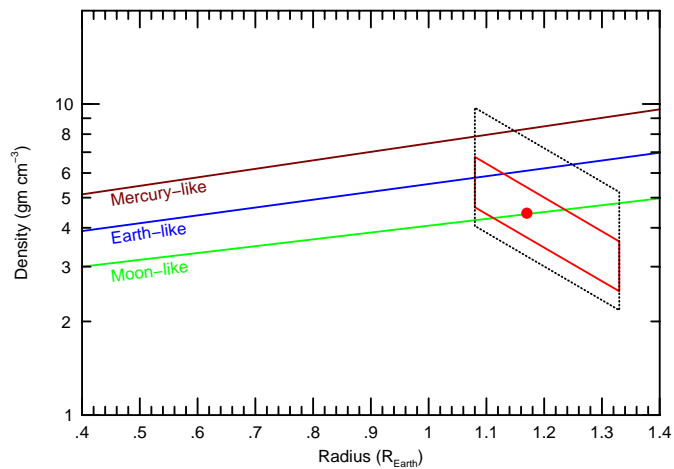}}
\caption{The density-radius relationships for terrestrial planets
having a structure that is  Mercury-like (top curve), 
Earth-like (middle curve) or Moon-like (bottom curve). The large dashed parallelogram marks roughly the range
of values from the various mass determinations for Kepler-78b. The point marks
are adopted solution and the error (solid parallelogram).
}
\label{density}
\end{figure}

Although the $K$-amplitude determination is good enough
to establish that Kepler-78b is a rocky planet, it is
insufficient to establish what {\it type} of rocky planet it is. Figure~\ref{density}
shows the density-radius relationship for terrestrial planets computed
by Diana Valencia and shown as Figure 11 in Hatzes et al. (2010). The 
dashed parallelogram shows the range of density and radius values for
Kepler-78b, including 1$\sigma$ errors, from the various mass determinations. For the 
planet radius and error we used the value of P2013. 
The range of possible densities for Kepler-78b encompasses
planets that are Mercury-like (i.e. large iron core) to Moon-like
(where we define ``Moon-like'' as having no iron  core).
Although most density determinations are consistent with
an Earth-like structure, the  nominal 
density from our adopted solution implies that Kepler-78b has
a structure more like the Moon, i.e. one with a small iron core.

\section{Discussion}

The RV detection of Kepler-78b is robust in spite of the $K$-amplitude
being less than the measurement error and at least a factor of
five smaller than the amplitude of the activity-related RV jitter. Regardless
of the approach one uses to filter out the activity, one arives at the same
value for the $K$-amplitude to within 1$\sigma$ (Fig.~\ref{models}). 

In particular,
the pre-whitening method seems to provide a good ``quick look'' method of finding the
amplitude of possible signals in RV data. 
Harmonic filtering where one fits the activity variations with
the rotation period and its harmonics is similar to pre-whitening, but with more restrictions
on the number and frequency of the components. It is thus not suprising that both
give comparable results. 

The Fourier amplitude spectrum of the Keck
data supports  the use of harmonic filtering as the rotational period and its first
3 harmonics are present. This is not the case with the HARPS-N data which
shows no clear evidence for either the adopted rotational period or its harmonics.
However, the derived RV amplitude seems to be relatively insensitive to the
details in how the activity signal is removed.

The RV data for Kepler-78b are so numerous and of such high quality that the planet
would have been detected without knowledge of the transit period. In fact, the
unfiltered Fourier amplitude spectrum alone indicates the presence of a planet signal
with a period of 0.355 d and a false alarm probability of $\sim$ 1\%. The planet
signal planet is significant in the RV data without filtering.

We have shown that the FCO periodogram can be an effective
tool for finding unknown short period, low-amplitude planets in RV data
that are dominated by activity-related RV variations. In these cases
classic periodograms probably would fail. Our simulations show
that the FCO periodogram could have discovered a 0.355-d period in the RV data even 
if the RV amplitude was as low as $K$ $\approx$ 1 m\,s$^{-1}$. This corresponds to a planet
mass of  1 $M_\oplus$.  This bodes well for finding low mass, ultra-short period planets even in the presence
of activity noise. 

When extracting the signal of very short period planets in the presence
of activity noise we prefer to use the FCO method. It is simple and makes
only one reasonable assumption that in the course of the
night the RV variations due to planetary orbital motion is much larger
than that due to rotational modulation by surface
active regions. One needs no knowledge of the rotational
period of the planet and we do not care if this changes due
to differential rotation. 
Additionaly, systematic 
errors on time scales of days or longer are also removed with FCO, as well as the influence of other
long period planets that may be in the system (Hatzes et al. 2010).
On the other hand,  harmonic analysis needs an accurate  rotation period, coupled with the
assumption that this does not change over the course of the observing season. Furthermore, there
is no way to deal with any nightly systematic errors with harmonic analysis or 
pre-whitening.

The FCO periodogram and offset fitting to the orbit only works, however,
if you 1) have good temporal sampling on a given night, and 2) a planetary orbital period much
less than the rotational period of the star which defines the characteristic
timescales of activity. Our simulations indicate that if the orbital period of the
planet is at least a factor of four shorter than the shortest period associated
with activity (the rotation period and its harmonics), the FCO periodogram should be
an effective tool for finding {unknown} short period planets in RV data and may give a better
measure of the $K$-amplitude.

When trying to detect RV variations due to low mass planets one may be tempted to just take
a few measurements at those times that correspond to the extrema of the RV curve (phases 0.25 and 0.75 for 
a transiting planet). After all, these should show the maximum RV variation from orbital
motion. However, in the presence of activity noise, or when the measurement error dominates
the $K$-amplitude of the planet, it may be wiser to simply invest more measurements
and sample more of the orbit. After all,  there is also information on the
$K$-amplitude in the shape of the RV.  In the case of Kepler-78b, the 
orbital solution  with all the HARPS-N
data yielded a larger $K$-amplitude than the fit to only the Keck data. 
If one were to only use the RV measurements taken at the expected extrema of the
RV curve one would derive an even larger $K$-amplitude (2.5 m\,s$^{-1}$).
A 
solution using only HARPS-N data  where more than two measurements were made per night produced
an amplitude consistent with the Keck result.
For Kepler-78b this difference is subtle, yet important
as it distinguishes whether it has a structure more like the Moon or Mercury.

We know that Kepler-78b is rocky planet, but we do not
know what  {\it kind} of rocky planet it is. The Doppler-determined planet
mass depends on the details in how the
activity signal is removed as well as having sufficient data. The various techniques produced
a planet  mass (including 1$\sigma$ errors) that ranged from as low  as 1.1 $M_{\oplus}$ to as  high
as 2.3 $M_{\oplus}$. These yielded bulk densities that were consistent with a Moon-like up to Mercury-like
internal structures. The data are insufficient to make the subtle distinction between
internal structures, but the precise type of rocky planet may hold the key to
understanding how these terrestrial planets form.

It would be possible to distinguish between the types of rocky planet
if we had an error of about 5\% on our
adopted $K$-amplitude.  We estimated the number of measurements
that would be required to achieve this accuracy 
by calculating orbital solutions using a subset of the RV data and sequentially 
adding more nights into the solution. We estimate that to get an error of 5\% on the $K$-amplitude would require
approximately 300 measurements with Keck-like sampling, or about 30 nights of observations.
This is a large investment in observing time, but it is important if we are to determine the true
structure of Kepler-78b.

Of course, even if we have an excellent measurement of the planet mass with low errors, the
density  error will be dominated by the errors in the stellar parameters. 
Fitting the transit light curve only yields the ratio of the panet to stellar radii. Likewise,
the velocity $K$-amplitude of the star due to the planet depends on the stellar mass. An accurate
planet density requires accurate stellar parameters.
This can only be accomplished with an asteroseismic determination
of the stellar radius. In this respect the PLATO mission (Rauer et al. 2014) will
provide a significant improvement in our understanding of the the internal structure of
rocky exoplanets. PLATO not only will search for transiting rocky planets around bright stars
for which characterization studies are possible, but it will also measure stellar
oscillations in order to determine asteroseismic values for the stellar mass and radius.

\begin{acknowledgements}
The author would like to thank Malcolm Fridlund for his reading of the manuscript and useful comments.
\end{acknowledgements}


\begin{thebibliography}{}

\bibitem[Hatzes et~al (2010)]{2010A&A...520A..93H} Hatzes, A.P., Dvorak, R., Wuchterl, G. et al. 2010, A\&A, 520, 93

\bibitem[Hatzes et~al (2011)]{2011ApJ...743...75H} Hatzes, A.P., Fridlund, M., Nachmani, G. et  al. 2011, A\&A, 743, 75


\bibitem[Hatzes (2013a)]{2013AN....334..616H} Hatzes, A.P. 2013a, AN, 334, 616

\bibitem[Hatzes (2013b)]{2013ApJ...770..133H} Hatzes, A.P. 2013b, ApJ, 770, 133


\bibitem[Howard et~al (2013)]{2013Natur.503..381H} Howard, A., Sanchis-Ojeda, R., Marcy, G.W. et al. 2013,
Nature, 503, 381 (H2013)

\bibitem[Jefferys et~al. (1988)] {1988CeMec..41...39J} Jefferys, W.H., Fitzpatrick, M.J., McArthur, B.E.,
Celestial Mechanics, 41, 39

\bibitem[Kuschnig et~al (1997)]{1997A&A...328..544K} Kuschnig, R., Weiss, W.W., Gruber, R., Bely, P.Y., Jenkner, H. 1997, 328,
544

\bibitem[Lomb (1976)]{1976Ap&SS..39..447L} Lomb, N.R. 1976, Ap\&SS, 39, 447

\bibitem[Pepe et~al (2013)]{2013Natur.503..377P} Pepe. F., Cameron, A.C., Latham, D.W. et al. 2013, Nature, 503,
377 (P2013)

\bibitem[Queloz et~al (2009)]{2009A&A...506..303Q} Queloz, D., Bouchy, D., Moutou, C. et al. 2009, A\&A, 506, 303

\bibitem[Rauer et~al. (2014)]{} Rauer, H., Catala, C., Aerts, C. et al. 2014, arXiv: 1300.0696, to appear
in Experimental Astronomy

\bibitem[Sanchis-Ojeda et~al (2013)]{2013ApJ...774...54S} Sanchis-Ojeda, R., Rappaport, S., Winn, J.N.,
Levine, A., Kotson, M.C., Latham, D.W., Buchhave, L. 2013, ApJ, 774, 54

\bibitem[Scargle (1982)]{1982ApJ...263..835S} Scargle, J.D. 1982, ApJ, 263, 835

\bibitem[Zechmeister \& K\"urster (2009)]{2009A&A...496..577Z} Zechmeister, M., K\"urster, M. 2009, A\&A, 496, 577


\end{thebibliography}
\end{document}